# Liquid-liquid equilibria for (2-hydroxy benzaldehyde + n-alkane) mixtures. Intermolecular and proximity effects in systems containing hydroxyl and aldehyde groups

JUAN ANTONIO GONZÁLEZ*[1], CRISTINA ALONSO-TRISTÁN[2], FERNANDO HEVIA[1], LUIS F. SANZ[1], AND ISAÍAS GARCÍA DE LA FUENTE[1]

[1] G.E.T.E.F., Departamento de Física Aplicada, Facultad de Ciencias, Universidad de Valladolid, Paseo de Belén, 7, 47011 Valladolid, Spain,

[2] Dpto. Ingeniería Electromecánica. Escuela Politécnica Superior. Avda. Cantabria s/n. 09006 Burgos, (Spain)

*e-mail: jagl@termo.uva.es; Fax: +34-983-423136; Tel: +34-983-423757

**Abstract**

The liquid-liquid equilibrium (LLE) curves have been determined for the {2-hydroxyl-benzaldehyde (salicylaldehyde, SAC) + $CH_3(CH_2)_nCH_3$} mixtures ($n$ = 5,6,7,8,9). The equilibrium temperatures were determined observing, by means of a laser scattering technique, the turbidity produced on cooling when a second phase takes place. All the systems show an upper critical solution temperature, which linearly increases with $n$. Intermolecular effects have been investigated in (alkanol + benzaldehyde) systems using data from the literature. Interactions in 1-alkanol mixtures are mainly of dipolar type. The corresponding excess molar enthalpies, $H^E_m$, are large and positive, which reveal that interactions between like molecules are dominant. Interactions between unlike molecules are stronger for the methanol-containing system. For the other mixtures, the enthalpy of the 1-alkanol-benzaldehyde interactions remains more or less constant. At 298.15 K and equimolar composition, the replacement of a linear polar solvent by the isomeric aromatic one leads to increased $H^E_m$ values in systems with a given 1-alkanol. The (phenol + benzaldehyde) system shows strongly negative deviations from the Raoult's law. Proximity effects have been examined in (SAC + hydrocarbon) mixtures. Alkane-containing systems are essentially characterized by dipolar interactions, while dispersive interactions are prevalent in the solution with benzene. All the mixtures have been treated in terms of DISQUAC. The interaction parameters for the OH/CHO contacts and for the SAC/aromatic and SAC/alkane contacts have been reported. DISQUAC provides a correct description of the thermodynamic properties considered. In the case of SAC systems, this is done by defining a new specific group HO―C═C―CHO for salicylaldehyde.

## 1. Introduction

When two or more polar groups are situated within the same linear molecule, interactions between such polar compounds are stronger than those between linear molecules containing only one of the mentioned groups. For example, $H_{m,eq}^{E}$ (hereafter, excess molar enthalpy at equimolar composition and 298.15 K) for the (1-butanol + heptane) mixture is 655 J·mol$^{-1}$ [1]. However the upper critical solution temperature (UCST) of the system formed by 2-methoxyethanol (2ME, isomeric molecule of 1-butanol) and heptane is 319.7 K [2]. This clearly demonstrates that 2ME-2ME interactions are stronger than those between 1-butanol molecules, which is typically ascribed to the so-called proximity effects, *i.e.*, to intramolecular effects existing between the several polar groups within the same molecule. Proximity effects are highly complex since depend on the number and nature of the groups present in the molecule and on the separation between them. Thus, ether-ether interactions become stronger in the sequence: dipropylether < 2,5-dioxahexane < 2,5,8-trioxanonane < 2,5,8,11-tetraoxadodecane < 2,5,8,11,14-pentaoxapentadecane. This is supported by the relative variation of the $H_{m,eq}^{E}$ values of the mixtures of each one of the mentioned ethers with heptane. In the same order that above, $H_{m,eq}^{E}$ /J·mol$^{-1}$ : 204 [3] < 1285 [4] < 1621 [5] < 1705 [6] < 1897 [7]. Similarly, interactions between 2-(2-methoxyethoxy)ethanol (22MEE) molecules are stronger than 2ME-2ME interactions as is shown by the greater UCST of the (22MEE + heptane) system (381.1 K [8]). It is remarkable that interactions become weaker when the distance between the polar groups increases along the series $\alpha,\varpi$-dichloroalkane [9], or $\alpha,\varpi$-dibromoalkane [10] + *n*-alkane. For example, $H_{m,eq}^{E}$ (heptane)/J·mol$^{-1}$ = 1840 (ClH$_2$C−CH$_2$Cl) [11,12] > 1280 (ClH$_2$C−(CH$_2$)$_3$−CH$_2$Cl) [13]. However, the proximity of two oxygen atoms in linear acetals weakens the molecular interactions with regards to those existing between polyethers of similar size [14].

Intermolecular effects come into play when the polar groups belong to different molecules, and interactions between unlike molecules must be also taken into account. The overall result depends on the nature of the involved groups. It is to be noted that interactions between unlike molecules are dominant in 1-alkanol + linear amine systems [15], which leads to very negative $H_{m,eq}^{E}$ values for these mixtures { $H_{m,eq}^{E}$ = − 3200 J·mol$^{-1}$ for the (methanol + hexylamine) system [16]}. In contrast, interactions between like molecules are still predominant in the solutions formed by a 1-alkanol and solvents such as nitromethane ( $H_{m,eq}^{E}$ = 1296 J·mol$^{-1}$ for the methanol-containing mixture [17]).

The replacement of a linear aliphatic chain by an aromatic ring with the same number of C atoms is responsible of enhanced dipolar interactions between the polar aromatic molecules (aromaticity effect). Thus, the UCST of aniline + heptane system is 343.1 K [18], while hexylamine is miscible with heptane at 298.15 K at any concentration and $_{m\,eq}$ /J·mol$^{-1}$ = 962

[19]. The attachment of a second polar group to the phenyl ring modifies the thermodynamic properties of the considered systems, which depend on the nature of the groups and on their relative position [20]. Mixtures including ethoxybenzeneamine, a molecule which contains the $-NH_2$ and the $-O-$ groups, and octane show UCST/K = 304.1 (2-ethoxybenzeneamine); 358.2 (4-ethoxybenzeneamine) [20]. On the other hand, proximity effects between the aromatic ring and the polar group are also present when the latter is placed within a linear chain attached to the ring, as it occurs, *e.g.*, in systems containing $C_6H_5-(CH_2)_nCl$ [21] or $C_6H_5-CO-(CH_2)_nCH_3$ [22,23] or $C_6H_5-(CH_2)_nCN$ [24] and a *n*-alkane. We have investigated proximity effects in systems with linear molecules (alkoxyethanols [25], amino-ketones [26], linear polyethers [27,28]), as well as the aromaticity effect and related proximity effects. Thus, we have provided LLE data for benzaldehyde [29] or acetophenone or 4-phenyl-2-butanone [22,23], or aromatic amines [30,31], nitriles [24], or alkanols [32,33], or phenetidine [20] + alkane mixtures. In addition, many systems of this type [23,24,29,34,35] have been treated in the framework of the DISQUAC group contribution model [36,37], providing the corresponding interaction parameters. The present work is concerned with the investigation of intermolecular effects in (1-alkanol or phenol + benzaldehyde) systems. Previously, we have studied solutions formed by methanol or ethanol and propanal [38]. Proximity effects are also investigated by means of mixtures involving 2-hydroxy-benzaldehyde (salicylaldehyde, SAC), an aromatic molecule where the groups $-OH$ and $-CHO$ are present. At this end, we report LLE data for (SAC + heptane, or + octane, or + nonane, or + decane, or + undecane) systems. Moreover, all the solutions are characterized using DISQUAC. No interactions parameters are available in the framework of the modified UNIFAC model for the systems studied [39,40].

Salicylaldehyde is an important intermediate in chemical industries. It is used for the coumarin synthesis, relevant in the manufacture of soaps, flavours and fragrances. Coumarin also exhibits antioxidant, analgesic, anti-inflammatory, and antimutagenic properties [41]. Salicylaldehyde is a precursor of aspirin and its Schiff bases, as salicylaldoxime, are used for the recovery of Cu and other metals [42]. Schiff bases have gained attention since they exhibit antimicrobial and fungicidal activities [43] and have pharmacological and biological applications [44,45].

## 2. Experimental

*2.1 Materials.* Information about the source, purity, water content and density ($\rho$) of the pure compounds used in the experimental part of this research is included in Table 1. The chemicals were used as received. Density results were obtained using a vibrating-tube densimeter and a sound analyser, Anton Paar model DSA-5000. The repeatability and the relative standard uncertainty of the $\rho$ values are, respectively, $5\times10^{-3}$ kg·m$^{-3}$, and 0.002. Our

$\rho$ results for *n*-alkanes are in good agreement with those reported in the literature (Table 1). A careful literature survey showed that no $\rho$ value at 298.15 K is available for 2-hydroxybenzaldehyde. The data listed in Table 1 at 303.15 K and 318.15 K are more or less consistent with our measurement at 298.15 K. Water contents of the compounds studies were determined by the Karl-Fischer method. The relative standard uncertainty of the corresponding measurements is estimated to be 0.02.

*2.2 Apparatus and Procedure*

Mixtures were prepared by mass in small Pyrex tubes (0.009 m i.d. and about 0.04 m length; free volume of the ampoule ≈1.17×10$^{-6}$ m$^3$). Weights were obtained from an analytical balance Sartorius NSU125p (weighing accuracy 10$^{-8}$ kg). The tubes were immediately sealed by capping at 0.1 MPa and 298.15 K. Mole fractions were calculated on the basis of the relative atomic mass Table of 2015 issued by the Commission on Isotopic Abundances and Atomic Weights (IUPAC) [46].

The LLE curves were determined by means of the observation of the turbidity produced on cooling (1.2 K·h$^{-1}$) when a second phase takes place. Details on the experimental method applied and on the apparatus calibration can be found elsewhere [20]. The equilibrium temperatures were measured using a Pt-1000 resistance. Two or three runs are conducted for a better assessment of these temperatures. The thermometer was calibrated according to the ITS-90 scale of temperature using the triple point of the water and the melting point of Ga as fixed points. The precision of the equilibrium temperature measurements is ± 0.001 K. The corresponding estimated standard uncertainty depends on the region where measurements are conducted. In the flat region of the LLE curves (top of the curves), the uncertainty of the temperature is 0.1 K. Outside from this region (tails of the curves), it is 0.2 K. For the equilibrium mole fractions, the standard uncertainty is 0.0005. This value is estimated taking into account that the more volatile component is partially evaporated to the mentioned free volume of the ampoule.

### 3. Experimental results

The directly measured liquid-liquid equilibrium temperatures, $T$, vs. $x_1$, the mole fraction of the (SAC + *n*-C$_7$, or + *n*-C$_8$, or + *n*-C$_9$, or + *n*-C$_{10}$, or + *n*-C$_{11}$) systems are collected in Table 2 (Figure 1). Our result for the UCST of the heptane solution (291.4 K) is much lower than that available in the literature (307.1 K) [47]. No information about the purity of the salicylaldehyde used is given in such reference, but probably the compound had a rather high water content since it is well known that this type of impurity noticeably increases the UCST

[48,49]. Some features of the LLE curves of the studied mixtures are: (i) they show a rather flat maximum (Figure 1); (ii) the curves become progressively skewed to higher $x_1$ values when alkane size increases (Figure 1); (iii) The UCST values increase more or less linearly with the number of C atoms of the *n*-alkane (Table 3). In the present research, UCST/K = $273.6 + 2.47n$, ($r^2$ = 0.998) where *n* is the number of C atoms of the *n*-alkane. The mentioned trends are also encountered in many other systems previously investigated [2,20,22-24,28-33].

The experimental ($x_1$, $T$) data of each system were correlated by means of the equation [50,51]:

$$T/K = T_c/K + k|y - y_c|^m \tag{1}$$

with

$$y = \frac{\alpha x_1}{1 + x_1(\alpha - 1)} \tag{2}$$

$$y_c = \frac{\alpha x_{1c}}{1 + x_{1c}(\alpha - 1)} \tag{3}$$

In equations (1-3), *m*, *k*, $\alpha$, $T_c$ and $x_{1c}$ are the parameters which must be fitted against the experimental data. The coordinates of the critical point are denoted by ($x_{1c}$, $T_c$). It is remarkable that, when $\alpha = 1$, equation (1) is similar to [52-54]:

$$\Delta\lambda = B\tau^\beta \tag{4}$$

In this equation, $\Delta\lambda_1 = \lambda_1' - \lambda_2''$ is any order parameter, that is, any density variable in the conjugate phase (along this research, $\lambda_1 = x_1$). On the other hand, $\tau$, defined as $\tau = (T_c - T)/T_c$, is the reduced temperature, $\beta$ the critical exponent related to $\Delta\lambda_1$ and *B* stands for the amplitude [54]. It is well-known that the critical exponent $\beta$ depends on the theory applied to its determination [52,55].

The parameters $m$, $k$, $\alpha$, $T_c$ and $x_{1c}$ were obtained from an adjustment based on a Marquardt algorithm [56] with all the points weighted equally. Final values of the parameters, together with the standard deviations for the liquid-liquid equilibrium temperatures, $\sigma(T)$, are given in Table 3. The $\sigma(T)$ values are calculated from the equation:

$$\left(\sigma(T)/K\right) = \left[\sum \left(T_{exp}/K - T_{calc}/K\right)^2 / (N-n)\right]^{1/2} \quad (5)$$

Here, $N$ equals the number of data points, and $n$, the number of adjusted parameters (= 5). Results listed in Table 3 show that equation (1) correctly fits the experimental results.

## 4. DISQUAC model

The group contribution model DISQUAC is based on the rigid lattice theory developed by Guggenheim [57]. We provide now some important features of the model. (i) The geometrical parameters of the considered molecules, total molecular volumes, $r_i$, surfaces, $q_i$, and the molecular surface fractions, $\alpha_{si}$, are calculated additively on the basis of the group volumes $R_G$ and surfaces $Q_G$ recommended by Bondi [58]. At this end, the volume $R_{CH4}$ and surface $Q_{CH4}$ of methane are taken arbitrarily as equal to 1 [59]. The geometrical parameters for the groups used along the work are listed in Table 4. (ii) The partition function is factorized into two terms. The excess functions are the result of two contributions: a dispersive (DIS) term related to the dispersive forces; and a quasichemical (QUAC) term arising from the anisotropy of the field forces created by the solution molecules. For $G_m^E$, a combinatorial term, $G_m^{E,COMB}$, represented by the Flory-Huggins equation [59,60] has to be also included. Thus,

$$G_m^E = G_m^{E,DIS} + G_m^{E,QUAC} + G_m^{E,COMB} \quad (6)$$

$$H_m^E = H_m^{E,DIS} + H_m^{E,QUAC} \quad (7)$$

(iii) It is assumed that the interaction parameters depend on the molecular structure of the mixture compounds; (iv) The same coordination number (= 4) is used for all the polar contacts. This is a crucial shortcoming of the model, and is partially removed assuming that the interaction parameters are dependent on the molecular structure of the mixture compounds. (v) It is also assumed that $V_m^E = 0$.

The equations needed to calculate the DIS and QUAC contributions to $G_m^E$ and $H_m^E$ are given elsewhere [15,37]. The temperature dependence of the interaction parameters is expressed in terms of the DIS and QUAC interchange coefficients [15,37], $C_{st,l}^{DIS}$; $C_{st,l}^{QUAC}$ where s ≠ t are two contact surfaces present in the mixture and $l$ = 1 (Gibbs energy; $C_{st,1}^{DIS/QUAC} = g_{st}^{DIS/QUAC}(T_o)/RT_o$); $l$ = 2 (enthalpy, $C_{st,2}^{DIS/QUAC} = h_{st}^{DIS/QUAC}(T_o)/RT_o$)), $l$ = 3 (heat capacity, $C_{st,3}^{DIS/QUAC} = c_{pst}^{DIS/QUAC}(T_o)/R$)). $T_o$ = 298.15 K is the scaling temperature and $R$, the gas constant. The equations are available elsewhere [15,37].

As in previous applications, the LLE curves were calculated using DISQUAC taking into account that the values of the mole fraction $x_1$ of component 1 ($x_1', x_1''$) relating to the two phases in equilibrium are such that the functions $G_m^{M'}, G_m^{M''}$ ($G_m^M = G_m^E + G_m^{ideal}$) have a common tangent [61].

### 5. Adjustment of DISQUAC parameters

In terms of DISQUAC, the studied systems are regarded as possessing the following types of surfaces: (i) type a, aliphatic ($CH_3$, $CH_2$, in $n$-alkanes, or 1-alkanols); (ii) type d (CHO in benzaldehyde); type h, (OH in 1-alkanols or phenol) (iii) type s (s = b, $C_6H_5$ in phenol, or benzaldehyde, or $C_4H_4$ in SAC; s = c-$CH_2$ in cyclohexane); (iv) type y (HO – C – C – CHO in SAC, see below).

The general procedure applied in the estimation of the interaction parameters have been explained in detail in earlier works [15,37]. Final values of our fitted parameters are listed in Table 5. Some important remarks are the following.

#### 5.1 Phenol + benzaldehyde

This mixture is characterized by three contacts: (b,d), (b,h) and (d,h). The former is represented by only DIS interaction parameters [29]. For the (b,h) contacts, the interaction parameters, both dispersive and quasichemical, are taken equal to those obtained for the phenol + benzene system [34]. The $C_{dh,l}^{DIS}$ and $C_{dh,l}^{QUAC}$ coefficients (Table 5) are determined from VLE and $H_m^E$ data available for the present mixture [62].

#### 5.2 1-alkanol + benzaldehyde

These systems are built by the contacts: (a,b), (a,d), (a,h), (b,d), (b,h), (d,h). The interaction parameters for the (a,b) contacts are dispersive and are already known from the investigation of (alkyl-benzene + alkane) mixtures [63]. The (a,d) contacts are described by

DIS and QUAC interaction parameters, and are known from the study of (benzaldehyde + *n*-alkane) mixtures [29]. The (a,h) and (b,h) contacts are also represented by DIS and QUAC parameters determined from (1-alkanol + *n*-alkane [64], or + toluene [65,66]) systems. Therefore, only those for the (d,h) contacts remain to be fitted. This has been done using the corresponding data from the literature [67].

### 5.3 Salicylaldehyde + hydrocarbon

Theoretical calculations on the basis of considering salicylaldehyde as a molecule formed by three surfaces (aromatic, hydroxyl, aldehyde) show that the thermodynamic properties of the involved mixtures are not properly described. Particularly, the LLE curves of SAC + *n*-alkane systems become skewed towards very low mole fractions of the polar component. The latter solutions are built by 4 surfaces (aliphatic, aromatic, hydroxyl, aldehyde) which generate six contacts. It is remarkable that the mentioned DISQUAC calculations on LLE strongly depend on the interaction parameters for the hydroxyl/aldehyde contacts. The poor results obtained indicate that, proceeding in such way, the interactions between the hydroxyl and aldehyde groups are not correctly represented and that some new effect, namely proximity effects, must be specifically taken into account. For this reason, we decided to define a new group "y", HO–C–C–CHO in salicylaldeyde, with unknown geometrical parameters. In such a case, the (SAC + benzene) system is built by only one contact (b,y) and the (SAC + alkane) mixtures are built by three contacts: (s,b), (b,y) and (s,y) with s = a in *n*-alkane solutions and s = c in the cyclohexane system. The interaction parameters of the (s,b) contacts are dispersive and are already known [63]. Due to the lack of experimental data, in this application we do not distinguish between the (a,y) and (c,y) contacts which are assumed to be described by the same interchange coefficients. This approach is acceptable since it is well known, in the framework of DISQUAC, that $CH_2$/X and c-$CH_2$/X contacts (where X is any polar group) are represented by the same QUAC interchange coefficients. Some examples are: X = ether [14], chlorine [9], carbonyl [23,68], OCOO [69] (linear organic carbonates), N-CO [70] (tertiary amides), or OH [71]. We note that the SAC + benzene system shows a nearly symmetrical $H_m^E$ curve with a maximum rather low (364 J.mol$^{-1}$) [72]. In addition, the corresponding excess heat capacity at constant pressure is slightly negative ( $-0.7$ J·mol$^{-1}$·K$^{-1}$ at $x_1$ =0.5 and 298.15 K) [72]. These are typical features of mixtures essentially characterized by dispersive interactions. In other words, the (b,y) contact can be assumed entirely dispersive. On the other hand, in view of such considerations, one can expect that the magnitude of these interaction parameters will be rather low. With this in mind, the geometrical parameters of SAC and the $C_{\text{ay},l}^{DIS/QUAC}$ ($l$ = 1,2,3) coefficients are simultaneously fitted against the thermodynamic properties of SAC + alkane mixtures assuming, in this first step, that $C_{\text{by},l}^{DIS}$ ($l$ =1,2,3) = 0. Using the geometrical parameters obtained in this way (Table 4), the interaction parameters for the (b,y) contacts are now determined (Table 5). Finally, with these values, the $C_{ay,l}^{DIS}$ ($l$ = 1,2,3) coefficients are recalculated, keeping without any modification along this process the $C_{ay,l}^{QUAC}$ ($l$ = 1,2,3) values previously determined (Table 5).

## 6. Discussion

### 6.1 Alkanol + benzaldehyde

It is well known that $H_m^E$ values of systems formed by 1-alkanol and a solvent containing a polar group X can be considered, if structural effects are neglected [52,73], as the result of three contributions. Two of them are positive and arise from the disruption of alkanol-alkanol and solvent-solvent interactions along the mixing process. We represent those contributions by $\Delta H_{\text{OH-OH}}$ and $\Delta H_{\text{X-X}}$, respectively. The third contribution, $\Delta H_{\text{OH-X}}$, is negative and is due to the formation of interactions between unlike molecules upon mixing. Therefore, we can write [74-77]:

$$H_m^E = \Delta H_{\text{OH-OH}} + \Delta H_{\text{X-X}} + \Delta H_{\text{OH-X}} \tag{8}$$

For (1-alkanol + benzaldehyde) systems, values of $H_m^E$ and of excess molar volumes, $V_m^E$, are large and positive. Thus, $H_{m,eq}^E$/J·mol$^{-1}$ = 986 (methanol); 1638 (1-propanol); 1962 (1-pentanol) [67] (Table 6); and $V_m^E$ ($x_1$ = 0.5, 298.15 K)/cm$^3$·mol$^{-1}$ = 0.741 (1-butanol); 0.819 (1-butanol) [78]. Therefore, the dominant contributions to the mentioned excess functions come from the breaking of interactions between like molecules. In addition, our previous studies on (1-alkanol + acetophenone [23], or + benzonitrile [24], or + nitrobenzene [79]) allow the statement that, in the corresponding benzaldehyde mixtures, dipolar interactions are also relevant. According to this fact, the shape of the $H_m^E$ curves of (1-alkanol + C$_6$H$_5$CHO) systems are nearly symmetrical [67] (Figure 2). The enthalpy related to the intermolecular effects between 1-alkanol and benzaldehyde, $\Delta H_{\text{OH-CHO}}$, can be evaluated by extending the equation (8) to $x_1 \to 0$ [74,77,80]. In such a case, the magnitudes $\Delta H_{\text{OH-OH}}$ and $\Delta H_{\text{CHO-CHO}}$ are replaced by $H_{m1}^{E,\infty}$

(partial excess molar enthalpy at infinite dilution of the first component) of (1-alkanol or benzaldehyde + heptane) systems. Thus,

$$\Delta H_{\text{OH-CHO}} = H_{\text{m1}}^{\text{E},\infty}(1-\text{alkanol} + \text{benzaldehyde})$$

$$-H_{\text{m1}}^{\text{E},\infty}(1-\text{alkanol} + \text{heptane}) - H_{\text{m1}}^{\text{E},\infty}(\text{benzaldehyde} + \text{heptane}) \qquad (9)$$

For (1-alkanol + $n$-alkane) systems, we have assumed along a number of applications that the $H_{\text{m1}}^{\text{E},\infty}$ value is independent of the alcohol [23,74,81,82], and that $H_{\text{m1}}^{\text{E},\infty}$ = 23.2 kJ·mol$^{-1}$ [83-85]. For the benzaldehyde + heptane mixture, and for (1-alkanol + benzaldehyde) systems, the $H_{\text{m1}}^{\text{E},\infty}$ values have been determined using the coefficients obtained for the fittings of the experimental $H_{\text{m}}^{\text{E}}$ data to Redlich-Kister expansions for the mentioned systems [67,86]. Results listed in Table 7 show that $\Delta H_{\text{OH-CHO}}$ is slightly lower for the solution involving methanol and that remains roughly constant for the other systems. That is, interactions between unlike molecules are stronger for the methanol-containing mixture. The observed increase of $H_{\text{m}}^{\text{E}}$ with the alcohol size can be explained assuming that: (i) the hydroxyl group is more sterically hindered in longer 1-alkanols and this leads to a less negative contribution to $H_{\text{m}}^{\text{E}}$ related to the interactions between unlike molecules; (ii) the positive contribution to $H_{\text{m}}^{\text{E}}$ from the breaking of interactions between benzaldehyde molecules increases with the aliphatic surface of the 1-alkanol.

The systems (1-alkanol + acetophenone, or + anisole, or + ethyl benzoate, or + benzonitrile, or + nitrobenzene, or + chlorobenzene, or + aniline) also show positive $H_{\text{m}}^{\text{E}}$ values (Figure 3). For example, $H_{\text{m,eq}}^{\text{E}}$ (1-propanol)/J·mol$^{-1}$ = 1807 (nitrobenzene) [87]; 1454 (benzonitrile) [88]; 1364 (acetophenone) [89]; 784 (chlorobenzene) [90], 783 (aniline) [91]. Therefore, the contributions to $H_{\text{m}}^{\text{E}}$ from the breaking of interactions between like molecules are also now prevalent over that related to 1-alkanol-solvent interactions. Interestingly, $H_{\text{m,eq}}^{\text{E}}$ of the methanol + aniline system is negative ($-170$ J·mol$^{-1}$ [92]). However, such value can be ascribed to the existence of strong structural effects in this solution. In fact, its excess molar volume, $V_{\text{m}}^{\text{E}}$, is extremely negative ($-0.899$ cm$^3$·mol$^{-1}$, at 298.15 K and $x_1$ = 0.5, [93]), and assuming ideal behaviour for the isobaric expansion coefficient, $\alpha_p$, and for the isothermal compressibility, $\kappa_T$, it is possible to calculate the excess molar internal energy at constant volume from $U_{Vm}^E = H_m^E - \frac{\alpha_p}{\kappa_T} TV_m^E$ [52,73]. At equimolar composition and 298.15 K, the result is: $U_{Vm}^E = 192$ J·mol$^{-1}$.

On the other hand, it is to be noted that the replacing of heptane by toluene in mixtures with a given 1-alkanol leads to increased $H_{m,eq}^E$ values. Thus, $H_{m,eq}^E$ (1-propanol)/J·mol$^{-1}$ = 979 (toluene) [94]; 597 (heptane) [95]. This indicates that toluene is a better breaker of the alcohol self-association than heptane. $H_{m,eq}^E$ values of the systems (1-alkanol + acetophenone, or + benzaldehyde, or + benzonitrile, or + nitrobenzene) are greater than those of the mixtures with toluene (Figure 3). This suggests that the mentioned aromatic molecules are better breakers of the alkanol-alkanol interactions than toluene. Nevertheless, one should take into account that 1-alkanols are also good breakers of the dipolar interactions between these solvent molecules. In fact, many (aromatic polar compound + alkane) systems show miscibility gaps, with upper critical solution temperatures (UCST) not far from 298.15 K. For example, UCST(decane)/K = 277.4 (acetophenone) [23]; 278.5 (benzaldehyde) [29]; 296 (nitrobenzene) [96]. The (aniline + heptane) mixture is a more extreme case since UCST = 343.1 K [18]. The (1-alkanol + chlorobenzene, or + aniline) show lower $H_{m,eq}^E$ values than the corresponding toluene solutions (Figure 3). In the case of aniline systems, it might be ascribed to the existence of stronger interactions between unlike molecules. This matter deserves a more detailed investigation currently undertaken.

Intermolecular effects between 1-alkanols and a linear aldehyde (propanal) are drastically different. In fact, $H_{m,eq}^E$ /J·mol$^{-1}$ values of mixtures with propanal are extremely negative: −8360 (methanol) [97]; −6988 (ethanol) [98]. The (1-alkanol + linear amine) systems show a similar behaviour and their $H_{m,eq}^E$ values are also large and negative. For 1-hexylamine mixtures, $H_{m,eq}^E$ /J·mol$^{-1}$ = −3200 (methanol); −2456 (1-octanol) [16]. It is clear that the $H_m^E$ values of these mixtures are largely determined by strong intermolecular effects between unlike molecules. Interestingly, many other 1-alkanol mixtures containing isomeric linear or aromatic polar solvents are characterized by positive $H_{m,eq}^E$ values. This is the case, *e.g.*, for (1-alkanol + nitrile, or + alkanone) systems. Certainly, a direct comparison between $H_{m,eq}^E$ results for 1-alkanol + linear or aromatic compound mixtures is difficult due to the lack of the required data. However, if one takes into account that for a given 1-alkanol, $H_{m,eq}^E$ values decreases with the increasing of the solvent size, then it is rather clear that $H_{m,eq}^E$ (linear polar component) < $H_{m,eq}^E$ (isomeric aromatic polar component). For example, $H_{m,eq}^E$ (1-

propanol)/J·mol$^{-1}$ = 1395 (2-propanone) [99]; 1259 (2-heptanone) [100], being the latter value lower than that given above for the acetophenone system. This may be ascribed to the positive contribution to $H^E_{m,eq}$ from the disruption of interactions between solvent molecules is larger when aromatic molecules are involved, which seems to be a general trend. Thus, the same trend is observed for (1-alkanol + hexylamine, or + aniline) mixtures.

The (phenol + benzaldehyde) mixture behaves similarly to the (1-alkanol + propanal) system. It shows large and negative $H^E_{m,eq}$ values at very high temperatures ($-1366$ J·mol$^{-1}$ at 413.15 K and 1824 kPa) and it is characterized by negative azeotropes [62]. Consequently, intermolecular effects are dominant by far. We also find a variety of different behaviours for systems containing two aromatic polar compounds. ($H^E_{m,eq}$ / J·mol$^{-1}$) values are very negative for the systems aniline + 3-hydroxytoluene ($-2051$ at $x_1 = 0.471$, $T$/K = 301.15) [101], {2-chlorophenol + aniline ($-7550$) [102], or + pyridine ($-9460$), or + quinoline ($-9850$) [103]}. Mixtures formed by aniline and chlorobenzene, or methoxybenzene, or nitrobenzene are characterized by positive ($H^E_{m,eq}$ /J·mol$^{-1}$) values: 968 [104], 350 and 544 [105], respectively. It seems that the hydroxyl group plays a decisive role in the formation of strong intermolecular effects between unlike aromatic polar molecules. This needs further experimental confirmation.

### 6.2  Salicylaldehyde + hydrocarbon

As it has been previously mentioned, interactions in the (SAC + benzene) system are mainly dispersive. LLE data reported in this work show that dipolar interactions between salicyladehyde molecules are prevalent in the mixtures studied. Accordingly, the (SAC + cyclohexane) mixture shows a large $H^E_{m,eq}$ value (1697 J·mol$^{-1}$), and the corresponding curve is symmetrical. [72]. These features can be ascribed to the existence of proximity effects between the hydroxyl and the aldehyde groups. It is remarkable that the molecular structure of salicylaldehyde has been investigated by means of gas-phase electron diffraction and ab initio calculations at the MP2(FC)/6-31G* level. The data indicate a planar equilibrium structure for the molecule and the existence of an intramolecular hydrogen bonding of similar strength to that present in 2-nitrophenol [106].

It is pertinent to compare UCST values of SAC systems with UCST results for other similar solutions. For example, in the case of mixtures including heptane, UCST/K = phenol 327.1 (phenol) [107]; 316.2 (2-nitrophenol) [47]; 291.4 (salicylaldehyde; (this work));  279 (2-chlorophenol) [47]. For hexadecane mixtures, we have UCST/K = 360.7 (phenol); 340.5 (2-methoxyphenol); 334.6 (2-nitrophenol) [108], 313.1 (SAC, extrapolated value). That is, dipolar interactions are stronger between phenol molecules than between molecules of the type 2-X-$C_6H_4OH$ (X = $NO_2$; Cl; CHO; $OCH_3$). This might be ascribed to the relative decrease of the hydroxyl surface when a second polar group is attached to the aromatic ring. However, steric effects are also very important since UCST values for heptane mixtures with 4-chlorophenol (340.2 K, [47]) or with 4-nitrophenol (373 K, [8]) are higher than the corresponding result for the phenol solution (see above).

### *6.3 DISQUAC results*

The model provides a good representation of the thermodynamic properties considered: $H_m^E$ of alkanol + benzaldehyde, systems (Table 6, Figures 2,4), coordinates of the azeotropes of the phenol + benzaldehyde mixture, and LLE of salicylaldehyde + *n*-alkane systems (Table 3, Figure 1). DISQUAC provides for the phenol + benzaldehyde system negative azeotropes at $T_{az}$ /K = 388; $x_{1az}$ = 0.742; $P_{az}$/kPa = 9.44 and at $T_{az}$ /K = 447; $x_{1az}$ = 0.628; $P_{az}$/kPa = 72.6. The experimental results are $T_{az}$ /K = 388; $x_{1az}$ = 0.715; $P_{az}$/kPa = 10; and $T_{az}$ /K = 447, $x_{1az}$ = 0.600; $P_{az}$/kPa = 73 [62]. It is remarkable that the weak temperature dependence of $H_m^E$ for the phenol solution is correctly described by DISQUAC (Table 6). Larger discrepancies are obtained for $H_m^E$ of the 1-hexanol + benzaldehyde mixture, but this is due to the experimental results largely deviate from those for the remainder (1-alkanol + benzaldehyde) systems (Table 6). The theoretical LLE curves for (SAC + *n*-alkane) mixtures are more rounded than the experimental ones (Figure 1). This is due to, at 298.15 K, the system temperature is close to the UCST, and mean field theories as DISQUAC, which assume that the excess molar Gibbs energy is an analytical function close to the critical point, cannot represent the flattening of the LLE curves at that condition [52,53]. Similar results have been obtained when investigating other mixtures at similar conditions [23,29,69,109,110]. Finally, we must underline two points. (i) the QUAC interaction parameters for the OH/CHO contacts in (1-alkanol + benzaldehyde) mixtures are independent of the 1-alkanol. The same occurs for (1-alkanol + alkanone [23,111], or *n*-alkanoate [112], or + linear organic carbonate [113]), with regards to the OH/X (X= CO, COO, OCOO) contacts for (phenol + *n*-alkane) [34]. Interestingly, different QUAC parameters are only encountered for the first members of homologous series for mixtures such as (1-alkanol + tertiary) amide [70,110], or + cyclic ether [114]. Thus, this seems to be a typical behaviour of the model. (ii) Good results are obtained for salicylaldehyde systems when defining the new HO – C – C – CHO group, which can be ascribed to the strong intramolecular effects existing in these solutions.

### 7. **Conclusions**

LLE measurements have been conducted for SAC + *n*-alkane mixtures, which are characterized by having an UCST. Intermolecular effects have been examined in alkanol +

benzaldehyde systems. Dipolar interactions are dominant in 1-alkanol mixtures, whose $H_m^E$ values are large and positive. Therefore, the main contribution to $H_m^E$ arises from the breaking of interactions between like molecules. 1-Alkanol-benzaldehyde interactions are stronger for the methanol mixture. For the remainder solutions, interactions between unlike molecules are more or less independent of the 1-alkanol. The phenol + benzaldehyde system shows strongly negative deviations from the Raoult's law. Proximity effects have been investigated by means of the SAC + alkane mixtures. All the systems have been treated in terms of DISQUAC. The model correctly describes the thermodynamic properties considered. In the case of SAC systems, this was done by defining a new specific group HO – C – C – CHO for salicylaldehyde.

TABLE 1

Chemicals abstract number, CAS, source, initial mole fraction, experimental (Exp.) and literature densities, $\rho$, and water contents expressed in mass fraction of pure compounds at 0.1 MPa and 298.15 K[a]

| Compound | CAS | Source | Initial mole fraction[b] | $\rho$ /kg·m$^{-3}$ Exp. | $\rho$ /kg·m$^{-3}$ Lit. | Water content/wt% |
|---|---|---|---|---|---|---|
| 2-hydroxy-benzaldehyde | 90-02-8 | Sigma-Aldrich | ≥0.994 | 1162.5 | 1148.3[115,c] | 56×10$^{-4}$ |
|  |  |  |  |  | 1141[116,d] |  |
| Heptane | 142-82-5 | Fluka | ≥0.995 | 679.71 | 679.46[117] | 12×10$^{-4}$ |
| Octane | 111-65-9 | Sigma-Aldrich | ≥0.994 | 698.84 | 698.62[117] | 15×10$^{-4}$ |
| Nonane | 111-84-2 | Fluka | ≥0.99 | 714.05 | 713.99[118] | 22×10$^{-4}$ |
| Decane | 124-18-5 | Fluka | ≥0.998 | 726.42 | 726.35[117] | 12×10$^{-4}$ |
| Undecane | 1120-21-4 | Fluka | ≥0.995 | 736.72 | 736.7[119] | 18×10$^{-4}$ |

[a]standard uncertainties are: $u(T) = 0.01$ K; $u(P) = 1$ kPa; the relative standard uncertainty for density $\rho$, is $u_\mathrm{r}(\rho) = 0.002$ and 0.02 for water content; [b]provided by the supplier from GC analysis; [c]value at 303.15 K; [d]value at 318.15 K

TABLE 2

Experimental liquid-liquid equilibrium temperatures vs. mole fraction, $x_1$, for 2-hydroxy-benzaldehyde (1) + n-alkane (2) mixtures[a] at 0.1 MPa.

| $x_1$ | T/K | $x_1$ | T/K |
|---|---|---|---|
| 2-hydroxy-benzaldehyde (1) + heptane (2) | | | |
| 0.1871[b] | 276.6 | 0.5815[c] | 291.2 |
| 0.2218[b] | 281.0 | 0.5862[c] | 291.1 |
| 0.2609[b] | 284.9 | 0.5868[c] | 291.1 |
| 0.3066[b] | 287.9 | 0.5911[c] | 291.1 |
| 0.3434[b] | 289.4 | 0.6242[c] | 290.8 |
| 0.3746[b] | 290.3 | 0.6576[c] | 290.0 |
| 0.4153[b] | 291.1 | 0.6764[c] | 289.3 |
| 0.4521[b] | 291.5 | 0.6970[c] | 288.6 |
| 0.4892[b] | 291.6 | 0.7301[c] | 286.2 |
| 0.5157[c] | 291.5 | 0.7578[c] | 283.9 |
| 0.5386[c] | 291.5 | 0.7879[c] | 280.0 |
| 0.5654[c] | 291.3 | 0.8179[c] | 275.6 |
| 2-hydroxy-benzaldehyde (1) + octane (2) | | | |
| 0.2498[b] | 282.4 | 0.5517[c] | 293.1 |
| 0.2783[b] | 285.1 | 0.5737[c] | 293.0 |
| 0.3025[b] | 286.7 | 0.5968[c] | 293.1 |
| 0.3230[b] | 288.2 | 0.6381[c] | 293.0 |

Table 2 (continued)

| | | | |
|---|---|---|---|
| 0.3458[b] | 289.7 | 0.6662[c] | 292.6 |
| 0.3604[b] | 290.2 | 0.7002[c] | 291.6 |
| 0.3825[b] | 291.0 | 0.7204[c] | 290.7 |
| 0.4026[b] | 291.8 | 0.7511[c] | 288.8 |
| 0.4440[b] | 292.8 | 0.7794[c] | 286.4 |
| 0.4762[b] | 293.0 | 0.8084[c] | 282.7 |
| 0.5176[c] | 293.1 | | |

2-hydroxy-benzaldehyde (1) + nonane (2)

| | | | |
|---|---|---|---|
| 0.2250[b] | 280.9 | 0.5968[c] | 295.9 |
| 0.2458[b] | 282.9 | 0.6002[c] | 295.9 |
| 0.2613[b] | 284.3 | 0.6284[c] | 295.9 |
| 0.2847[b] | 286.2 | 0.6600[c] | 295.8 |
| 0.3093[b] | 288.6 | 0.6874[c] | 295.6 |
| 0.3300[b] | 289.8 | 0.7191[c] | 295.3 |
| 0.3577[b] | 291.1 | 0.7488[c] | 294.5 |
| 0.3917[b] | 292.7 | 0.7727[c] | 293.0 |
| 0.4302[b] | 294.2 | 0.7987[c] | 291.2 |
| 0.4625[b] | 295.1 | 0.8125[c] | 290.0 |
| 0.4978[b] | 295.6 | 0.8251[c] | 288.7 |
| 0.5393[c] | 295.8 | 0.8448[c] | 286.5 |
| 0.5668[c] | 295.9 | | |

Table 2 (continued)

2-hydroxy-benzaldehyde (1) + decane (2)

| | | | |
|---|---|---|---|
| 0.2369[b] | 281.9 | 0.6159[c] | 298.1 |
| 0.2593[b] | 283.8 | 0.6468[c] | 298.0 |
| 0.2942[b] | 286.9 | 0.6482[c] | 298.1 |
| 0.3185[b] | 289.0 | 0.6841[c] | 297.7 |
| 0.3325[b] | 289.9 | 0.7060[c] | 297.6 |
| 0.3694[b] | 292.9 | 0.7356[c] | 297.1 |
| 0.4043[b] | 294.4 | 0.7589[c] | 296.4 |
| 0.4471[b] | 296.3 | 0.7825[c] | 295.1 |
| 0.4826[b] | 297.1 | 0.8097[c] | 292.9 |
| 0.5185[c] | 297.6 | 0.8338[c] | 289.9 |
| 0.5570[c] | 297.9 | 0.8568[c] | 286.7 |
| 0.5919[c] | 298.1 | 0.8804[c] | 281.1 |

2-hydroxy-benzaldehyde (1) + undecane (2)

| | | | |
|---|---|---|---|
| 0.2990[b] | 287.6 | 0.6450[c] | 301.0 |
| 0.3374[b] | 290.4 | 0.6705[c] | 301.0 |
| 0.3462[b] | 291.7 | 0.7027[c] | 301.0 |
| 0.3808[b] | 294.3 | 0.7280[c] | 300.7 |
| 0.4028[b] | 295.7 | 0.7548[c] | 300.2 |
| 0.4307[b] | 297.1 | 0.7774[c] | 299.5 |

Table 2 (continued)

| | | | |
|---|---|---|---|
| 0.4527[b] | 298.3 | 0.7979[c] | 298.5 |
| 0.4824[b] | 299.3 | 0.8222[c] | 296.8 |
| 0.5131[c] | 299.9 | 0.8357[c] | 295.5 |
| 0.5436[c] | 300.5 | 0.8550[c] | 293.0 |
| 0.5747[c] | 300.7 | 0.8660[c] | 291.5 |
| 0.5859[c] | 300.9 | 0.8893[c] | 287.0 |
| 0.6103[c] | 301.1 | | |

[a]standard uncertainties are: $u(x_1) = 0.0005$; $u(p) = 1$ kPa; the combined expanded uncertainty (0.95 level of confidence) for temperature is $U_c(T) = 0.2$ K in the flat region of the curves and 0.4 K outside this region; [b]alkane rich phase; [c]salicylaldehyde rich phase

TABLE 3

Coefficients in eq. (1) for the fitting of the ($x_1$, $T$) pairs listed in Table 2 for 2-hydroxy-benzaldehyde (1) + $n$-alkane (2) mixtures; $\sigma(T)$ is the standard deviation defined by eq. (5). Coordinates of the critical points determined from the DISQUAC model using interchange coefficients listed in Table 5 are given between parentheses,

| $N$[a] | $m$ | $k$ | $\alpha$ | $T_c$/K | $x_{1c}$ | $\sigma(T)$/K |
|---|---|---|---|---|---|---|
| | | | 2-hydroxy-benzaldehyde (1) + heptane (2) | | | |
| 24 | 3.213 | − 627 | 0.886 | 291.4 | 0.511 | 0.13 |
| | | | | (292.7) | (0.498) | |
| | | | 2-hydroxy-benzaldehyde (1) + octane (2) | | | |
| 21 | 3.171 | − 604 | 0.721 | 293.1 | 0.556 | 0.10 |
| | | | | (294.2) | (0.533) | |
| | | | 2-hydroxy-benzaldehyde (1) + nonane (2) | | | |
| 25 | 3.069 | − 441 | 0.614 | 295.5 | 0.603 | 0.11 |
| | | | | (296.0) | (0.617) | |
| | | | 2-hydroxy-benzaldehyde (1) + decane (2) | | | |
| 24 | 3.220 | − 596 | 0.546 | 298.0 | 0.620 | 0.11 |
| | | | | (299.0) | (0.640) | |
| | | | 2-hydroxy-benzaldehyde (1) + undecane (2) | | | |
| 26 | 3.571 | − 523 | 0.529 | 301.0 | 0.649 | 0.07 |
| | | | | (304.6) | (0.677) | |

[a] number of experimental data points

TABLE 4

Relative group increments for molecular volumes, $r_G = V_G / V_{CH_4}$ and areas, $q_G = A_G / A_{CH_4}$ calculated from Bondi's method ($V_{CH_4}$ = 17.12×10$^{-6}$ m$^3$·mol$^{-1}$, $A_{CH_4}$ = 2.90×10$^{-5}$ m$^2$·mol$^{-1}$) or determined in this work

| Group | $r_G$ | $q_G$ | Ref. |
|---|---|---|---|
| CH$_3$– | 0.79848 | 0.73013 | [59] |
| –CH$_2$– | 0.59755 | 0.46552 | [59] |
| C$_6$H$_5$– | 2.67752 | 1.83797 | [59] |
| C$_4$H$_4$– | 1.8832 | 1.3816 | this work |
| –OH | 0.46963 | 0.50345 | [64] |
| –CHO | 0.88435 | 0.81724 | [120] |
| HO–C–C–CHO | 1.6775 | 1.0422 | this work |

TABLE 5

Dispersive (DIS) and quasichemical (QUAC) interchange coefficients, $C_{st,1}^{DIS}$ and $C_{st,1}^{QUAC}$, for (s,t) contacts[a] in binary mixtures containing the hydroxyl and the aldehyde groups or the salicylic group.

| Compound | $C_{st,1}^{DIS}$ | $C_{st,2}^{DIS}$ | $C_{st,3}^{DIS}$ | $C_{st,1}^{QUAC}$ | $C_{st,2}^{QUAC}$ | $C_{st,3}^{QUAC}$ |
|---|---|---|---|---|---|---|
| 1-alkanol + benzaldehyde; contact (s,t) = (d,h) | | | | | | |
| Methanol | −10[b] | −4.4 | | 3.5 | −3 | |
| Ethanol | −12[b] | −7.3 | | 3.5 | −3 | |
| 1-Propanol | −14[b] | −9.9 | | 3.5 | −3 | |
| 1-Butanol | −16[b] | −12.6 | | 3.5 | −3 | |
| 1-Pentanol | −16[b] | −16.1 | | 3.5 | −3 | |
| 1-Hexanol | −16[b] | −24 | | 3.5 | −3 | |
| Phenol + benzaldehyde; contact (s,t) = (d,h) | | | | | | |
| | −8.2 | 3 | −40 | 2 | −26 | −60 |
| Salicyladehyde + benzene; contact (s,t) = (b,y) | | | | | | |
| | 1[b] | 1.45 | | | | |
| Salicyladehyde + alkane; contact (s,t) = (a,y) | | | | | | |
| Heptane | −1.175 | 0.5 | | 3.5 | 2.2 | |
| Octane | −1.23 | 0.5 | | 3.5 | 2.2 | |
| Nonane | −1.27 | 0.5 | | 3.5 | 2.2 | |
| ≥ Decane | −1.295 | 0.5 | | 3.5 | 2.2 | |

[a] s ≠ t = a ($CH_3$; $CH_2$); b ($C_6H_6$, $C_6H_5$, $C_4H_4$); d (CHO); h (OH), y (HO−C−C−CHO). No distinction is made between aliphatic and cyclic groups in *n*-alkanes or cyclohexane in SAC mixtures. [b] guessed value

TABLE 6

Molar excess enthalpies, $H_{m,eq}^{E}$, at equimolar composition and temperature $T$ for alkanol + benzaldehyde systems or for salicylaldehyde + hydrocarbon mixtures.

| System | $T$/K | $N^a$ | $H_{m,eq}^{E}$ /J·mol$^{-1}$ | | $dev(H_m^E)^b$ | | Ref. |
|---|---|---|---|---|---|---|---|
| | | | Exp. | DQ | Exp. | DQ | |
| Methanol + benzaldehyde | 298.15 | 12 | 986 | 961 | 0.006 | 0.044 | [67] |
| ethanol + benzaldehyde | 298.15 | 12 | 1298 | 1289 | 0.003 | 0.028 | [67] |
| 1-propanol + benzaldehyde | 298.15 | 12 | 1638 | 1649 | 0.003 | 0.016 | [67] |
| 1-butanol + benzaldehyde | 298.15 | 12 | 1878 | 1883 | 0.005 | 0.009 | [67] |
| 1-pentanol + benzaldehyde | 298.15 | 12 | 1962 | 1926 | 0.005 | 0.017 | [67] |
| 1-hexanol + benzaldehyde | 298.15 | 9 | 1686 | 1648 | 0.013 | 0.021 | [121] |
| Phenol + benzaldehyde$^c$ | 363.15 | 15 | $-1405$ | $-1397$ | 0.004 | 0.018 | [62] |
| | 413.15 | 15 | $-1368$ | $-1335$ | 0.012 | 0.058 | [62] |
| Salicylaldehyde + benzene | 298.15 | 22 | 366 | 371 | 0.008 | 0.033 | [72] |
| Salicylaldehyde + cyclohexane | 298.15 | 13 | 1698 | 1730 | 0.004 | 0.019 | [72] |

$^a$number of experimental data points; $^b$ $dev(H_m^E) = \{\frac{1}{N}\sum\left[\frac{H_{m,exp}^E - H_{m,calc}^E}{H_{m,exp}^E(x_1 = 0.5)}\right]^2\}^{1/2}$ ; $^c$ $p = 1824$ kPa

TABLE 7

Partial molar excess enthalpies,[a] $H_1^{E,\infty}$, at $T = 298.15$ K, atmospheric pressure and at infinite dilution of the solute for solute(1) + organic solvent(2) mixtures, and hydrogen bond enthalpies, $\Delta H_{\text{OH-CHO}}$, for 1-alkanol (1) + benzylaldehyde (2) systems.

| System | $H_1^{E,\infty}$ /kJ·mol$^{-1}$ | $\Delta H_{\text{OH-CHO}}$ /kJ·mol$^{-1}$ |
|---|---|---|
| Benzylaldehyde (1) + heptane (2) | 9.1 [86] | |
| 1-alkanol (1) + heptane (2) | 23.2 [83-85] | |
| Methanol (1) + benzylaldehyde (2) | 5.4 [67] | −26.9 |
| Ethanol (1) + benzylaldehyde (2) | 8.1 [67] | −24.2 |
| 1-propanol (1) + benzylaldehyde (2) | 8.8 [67] | −23.5 |
| 1-butanol (1) + benzylaldehyde (2) | 8.8 [67] | −23.5 |
| 1-pentanol (1) + benzylaldehyde (2) | 7.6 [67] | −24.7 |

[a]values obtained from $H_m^E$ data over the whole concentration range

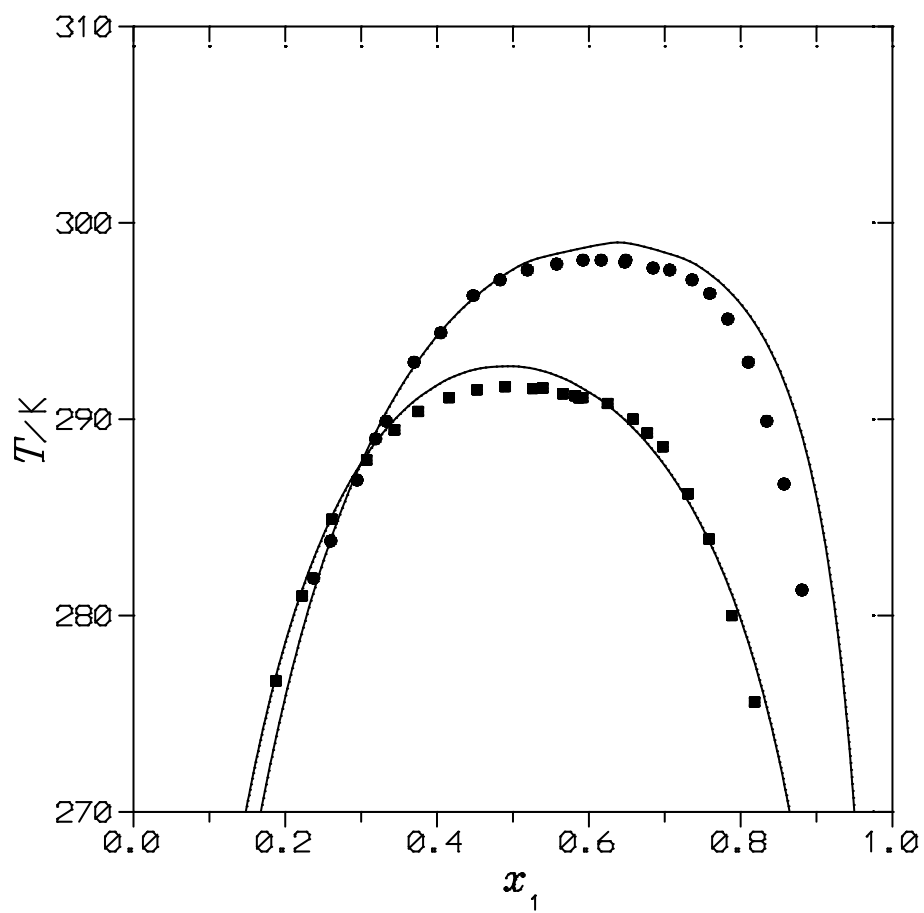

Figure 1    Liquid-liquid equilibrium temperatures, $T$, vs. $x_1$, the mole fraction of salicylaldehyde for salicylaldehyde (1) + $n$-alkane(2) mixtures. Points, experimental results (this work): (■), heptane; (●), decane. Solid lines, DISQUAC calculations using interaction parameters from Table 5.

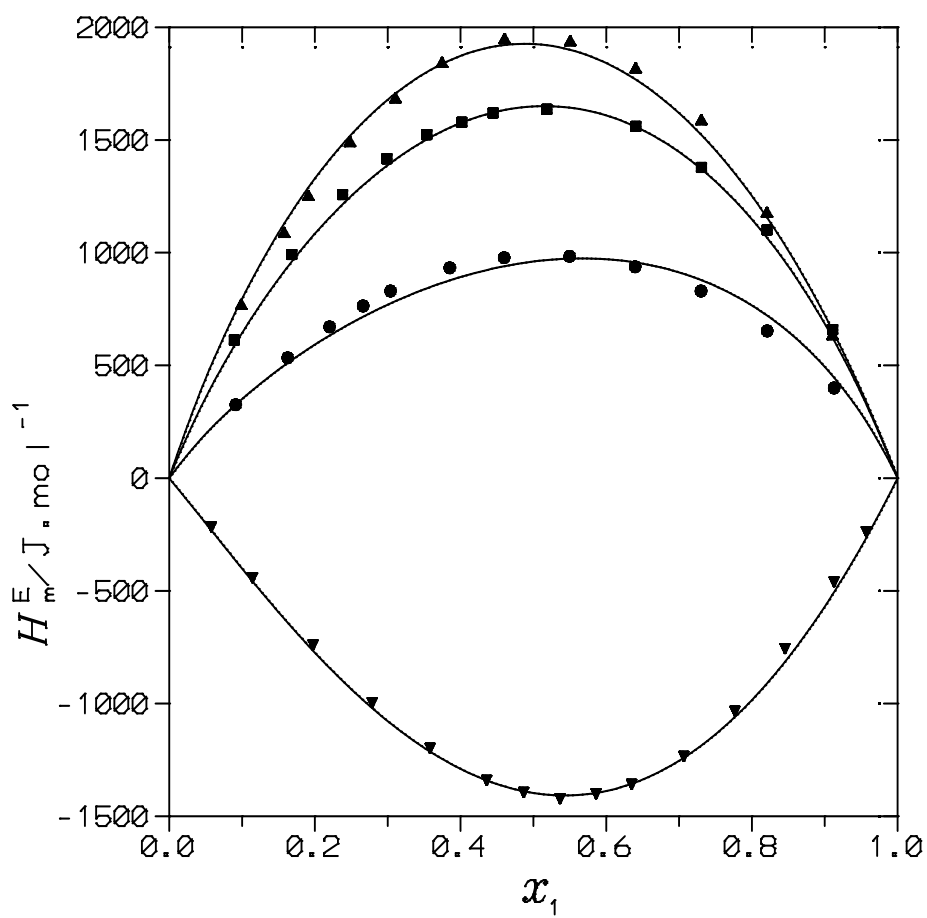

Figure 2    $H_m^E$ for alkanol (1) + benzaldehyde (2) mixtures vs $x_1$, the mole fraction of alkanol. Points, experimental results: (●), methanol; (■), 1-propanol; (▲), 1-pentanol ($T$ = 298.15 K, $p$ = 101.325 kPa [67]); (▼), phenol ($T$ = 363.15 K, $p$ = 1824 kPa [62]). Solid lines, DISQUAC calculations using interaction parameters from Table 5.

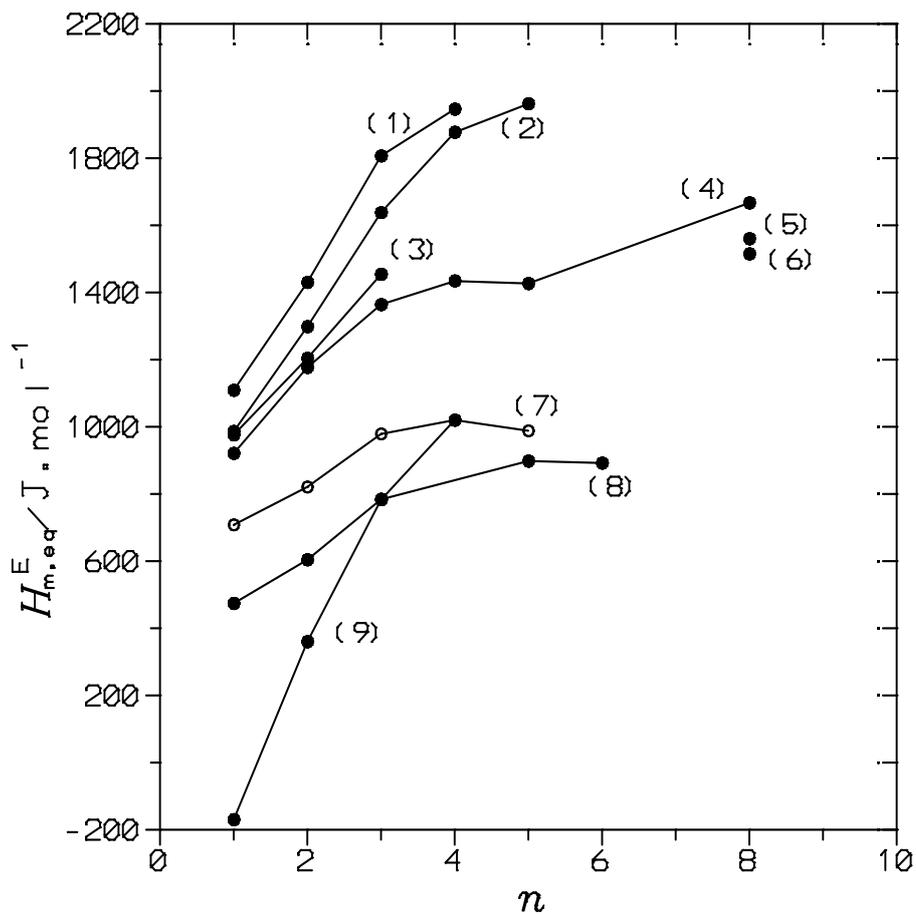

Figure 3    $H_{m.eq}^{E}$ for 1-alkanol(1) + organic solvent(2) mixtures at 298.15 K vs. *n*, the number of C atoms in 1-alkanols. Points, experimental results: (1), nitrobenzene [87]; (2), benzaldehyde [67]; (3), benzonitrile [88]; (4), acetophenone [89,122,123]; (5) anisole; (6), ethylbenzoate [123]; (7), toluene [94]; (8), chlorobenzene [89], (9), aniline [90-92,124,125]. The lines are only for the aid of the eye.

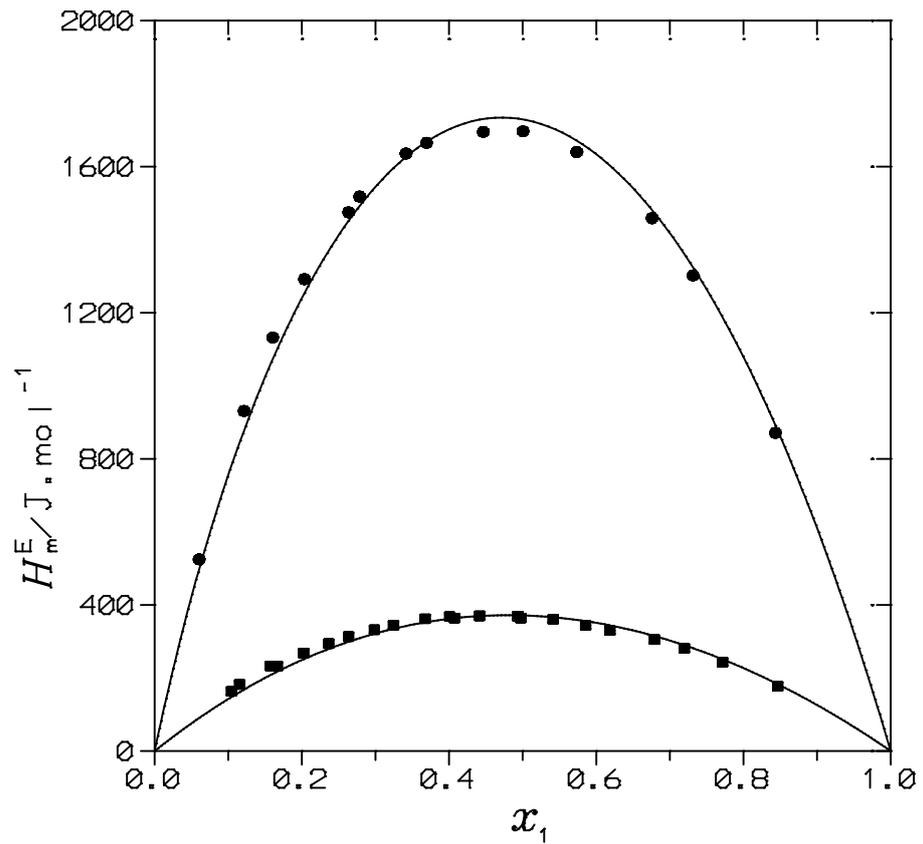

Figure 4    $H_m^E$ for salicylaldehyde(1) + hydrocarbon(2) mixtures vs $x_1$, the mole fraction of salicylaldehyde. Points, experimental results at 298.15 K and 0.1 MPa [72]: (■), benzene; (●), cyclohexane. Solid lines, DISQUAC calculations using interaction parameters from Table 5.